\documentclass[entropy,article,accept,pdftex,moreauthors,amsmath]{Definitions/mdpi} 
\usepackage{bm}
\newcommand{\average}[1]{\ensuremath{\langle#1\rangle}}

\newcommand{\ttau}{ \tilde{\tau} }
\newcommand{\teta}{ \tilde{\eta} }

\newcommand{\bk}{ \bm{k} }
\newcommand{\bxp}{ \bm{x}_\perp }
\newcommand{\refeq}[1]{(\ref{#1})}
\newcommand{\xm}{ x^- }
\newcommand{\xp}{ x^+ }
\newcommand{\xmp}{ x^\mp }
\newcommand{\xc}{ x_{\rm c} }
\newcommand{\qs}{ Q_{\rm s} }

\firstpage{1} 
\pubvolume{1}
\issuenum{1}
\articlenumber{0}
\pubyear{2023}
\copyrightyear{2023}
\datereceived{ } 
\dateaccepted{ } 
\datepublished{ } 
\hreflink{https://doi.org/} 

\Title{Effect of Longitudinal Fluctuations of 3D Weizs\"{a}cker-Williams Field on Pressure Isotropization of Glasma}

\TitleCitation{Effect of Longitudinal Fluctuations of 3D Weizs\"{a}cker-Williams Field on Pressure Isotropization of Glasma}

\Author{Hidefumi Matsuda $^{1}$*\orcidA{} and Xu-Guang Huang $^{1,2,3}$*\orcidB{}}

\AuthorNames{Hidefumi Matsuda and Xu-Guang Huang}

\AuthorCitation{Matsuda, H.; Huang, X.-G.}

\address{%
$^{1}$ \quad Physics Department and Center for Particle Physics and Field Theory, Fudan University, Shanghai 200438, China\\
$^{2}$ \quad Key Laboratory of Nuclear Physics and Ion-beam Application (MOE), Fudan University, Shanghai 200433, China\\
$^{3}$ \quad Shanghai Research Center for Theoretical Nuclear Physics, National Natrual Science Foundation of China and Fudan University, Shanghai 200438, China
}

\corres{Correspondence: da.matsu.00.bbb.kobe@gmail.com (H.M.); huangxuguang@fudan.edu.cn (X.-G.H.)}

\abstract{
We investigate the effects of boost invariance breaking on the isotropization of pressure in the glasma, using the $3+1$D glasma simulation.
The breaking is attributed to spatial fluctuations of the classical color charge density along the collision axis.
We present numerical results for pressure and energy density at mid-rapidity and across a wider rapidity region. 
It is found that, despite varying longitudinal correlation lengths, the behaviors of the pressure isotropizations are qualitatively similar. 
The numerical results suggest that, in the initial stage, longitudinal color electromagnetic fields develop, similar to those in the boost invariant glasma. 
Subsequently, these fields evolve into a dilute glasma, expanding longitudinally in a manner akin to a dilute gas.
We also show that the energy density at mid-rapidity exhibits a $1/\tau$ decay in the dilute glasma stage.
}

\keyword{Relativistic heavy-ion collisions; classical Yang-Mills theory; glasma} 

\begin{document}


\section{Introduction}
Relativistic heavy-ion collision experiments in nuclear physics aim to study various states governed by quantum chromodynamics (QCD), which describes the strong interaction within the standard model of particle physics. 
The collision deposits matter at extremely high energy and/or density in the collision region. 
Subsequently, this matter is believed to evolve into quark--gluon plasma (QGP), where quarks and gluons, normally confined within protons and neutrons in vacuum, are free to move independently. 
Recently, research has shifted from merely confirming the creation of QGP to exploring its properties through experiments. 
This includes investigating the effects of magnetic fields~\cite{B1,B2,B3,B4,B5,B6,B7,B8,B9,B10,B11,B12,B13,B14,B15,B16,B17,B18} and rotation in the QGP medium~\cite{omega1,omega2,omega3,omega4}, as well as the observability of novel phenomena related to quantum anomalies, such as the chiral magnetic effect~\cite{B6,B7} and chiral vortical effect~\cite{CVE1,CVE2,CVE3,CVE4}. 
We would like to emphasize that heavy-ion collision experiments provide not only the QGP but also other rich physics landscapes of QCD, including the non-equilibrium state before the formation of QGP and the phase transition from the QGP to the hadron gas.

The pre-equilibrium process between the non-equilibrium state initially produced by the collision and the QGP following relativistic hydrodynamics remains unclear. 
Since it is not trivial whether the initial matter with high energy density reaches the QGP before undergoing the QCD phase transition, due to the decrease in energy density caused by the rapid expansion of the created matter, a deeper understanding of the pre-equilibrium process reinforces our confidence that QGP is produced in experiments. 
Not only that, it is also vital for explaining the experimental results. 
For instance, improved modeling of the pre-equilibrium process can provide more accurate initial conditions for the hydrodynamic simulations used in the analyses of the experimental data. 
Additionally, photons emitted from the pre-equilibrium matter may carry information about the non-equilibrium state, as they are less affected by the medium. 
Of course, the non-equilibrium state of QCD matter itself is also an interesting subject of study.

In the study of non-equilibrium processes in heavy-ion collisions, a key issue is understanding how the created matter eventually behaves as a fluid governed by hydrodynamics. 
If the system were confined within a box, it might be expected to eventually reach equilibrium over time due to the interactions among the quarks and gluons. 
However, in reality the system expands at nearly the speed of light. 
Therefore, it is not an obvious question whether or not the created matter can even approach a hydrodynamic state in the first place.

Tracking the real-time evolution of the non-equilibrium state of QCD matter based on first-principle calculation is not practical with our existing computational resources. 
Therefore, several effective descriptions have been employed to reveal its qualitative characteristics.
For example, for the early stage of the collision, the classical Yang--Mills (CYM) field is an excellent description of the system, as explained in the next paragraph.

According to perturbative QCD, in a high-energy collision a nucleus can be considered as a dense state consisting of soft gluons.
Here, `soft' implies that the momentum carried is only a small fraction of the hadron's total momentum.
The color glass condensate (CGC) effective theory is a suitable description of such a high-energy nucleus~\cite{CGC}. 
In this theory, a high-energy nucleus is divided into soft partons (gluons) and hard partons (gluons and quarks with higher momentum than the soft gluons)~\cite{MV1,MV2,MV3}.
Soft gluons 
 are described as the CYM fields emitted from hard partons expressed as classical color charge density.
The name `color glass condensate' has its origins as follows: 
`Color' in QCD is a property analogous to electric charge. 
`Glass' is a disordered system and appears static in natural time. 
Similarly, hard partons are `frozen' in time due to the time dilation effects of special relativity, compared to the lifetime of soft gluons, yet can be rearranged by interactions. 
`Condensate' refers to the dense and coherent soft gluons, which are treated as the classical field, akin to a Bose--Einstein condensate. 
The theoretical description of the CGC is analogous to the Weizs\"{a}cker--Williams approximation in electromagnetic dynamics.
As a result of the collision between two CGCs, the initial state of the matter is expected to be a dense matter of soft gluons, a state referred to as `glasma'.
The glasma is well described by the CYM field, which is determined by solving the CYM equations of motion with \mbox{two classical} color sources, each representing the hard partons within different nuclei~\cite{glasma}.


Under the assumption of the infinite momentum limit, the glasma exhibits exact invariance under the boost transformation along the collision axis and, just after the collision, the color electromagnetic fields of such a boost invariant glasma are aligned along the collision axis~\cite{2D1,2D2,2D3,2D4,2D5,2D6,2D7,2D8}. 
However, in real relativistic heavy-ion collisions, there should be fluctuations that break the boost invariance. 
The CYM equation of motion has instabilities that lead to the rapid growth of these fluctuations~\cite{Instability1,Instability2,Instability3,Instability4,Instability5,Instability6,Instability7,Instability8,Instability9,Instability10,Instability11,Instability12,Instability13,Instability14,Instability15,Instability16,Instability17,Instability18,Instability19,Instability20,Instability21,Instability22}. 
Interestingly, these instabilities can drive pressure isotropization~\cite{Instability14,Instability15,Instability16,Instability17,Instability18,Instability19,Instability20,Instability21,Pressure} and entropy generation~\cite{Entropy1,Entropy2,Entropy3}. 

Here, a natural question arises regarding the extent to which the system approaches the hydrodynamic regime during the glasma stage.
Many studies on this issue focus on the isotropization of its pressure because sufficient isotropy in the pressure is crucial for the system to obey hydrodynamics. 
As mentioned above, the pressure isotropy is driven by the growth of fluctuations that break boost invariance, due to instabilities of the CYM theory. 
However, previous studies have not demonstrated the pressure isotropy consistent with hydrodynamic fluid.
For instance, some studies~\cite{Instability14,Instability15,Instability16,Instability17,Instability18,Instability19,Instability20,Instability21,Attractor1,Attractor2} have simulated the semiclassical time evolution of the glasma. 
These include fluctuations that break boost invariance due to minimal quantum uncertainty. 
They use classical statistical approximation (CSA) in a regime with much smaller coupling constants than realistic coupling constants. 
However, these simulations have not achieved sufficient pressure isotropy.
Another study~\cite{3DJIMWLK}, using the JIMWLK equation---which is the evolution equation for the separation scale between soft and hard partons in the CGC effective theory---incorporates the initial condition of the glasma that breaks boost invariance. 
This initial condition is derived from the momentum distribution of soft gluons in a nucleus with finite momentum. 
These studies also do not show sufficient pressure isotropy.
However, before concluding that the glasma does not exhibit the pressure isotropization under the classical approximation, it remains valuable to consider the effects originating from other physical mechanisms that break boost invariance.

In this study, we investigate the impact of boost invariance breaking, caused by spatial fluctuations of the classical color charge density along the collision axis, on the isotropization of pressure. 
This is done using the 3+1 dimensional glasma simulation without the boost invariance assumption, which has been studied recently~\cite{3Dglasma1,3Dglasma2,3Dglasma3,3Dglasma4,3Dglasma5,3Dglasma6}.
Boost invariance typically stems from the eikonal approximation for the classical color charge density, which is valid in the infinite momentum limit. 
The eikonal approximation includes three treatments of the classical color charge density or current: 
firstly, considering only the dominant component of the classical color current in the infinite momentum limit; 
secondly,~treating the classical color charge density as infinitely thin along the collision axis due to Lorentz contraction; 
and thirdly, assuming the classical color charge density is not changed by the collision.
However, in the 3+1D glasma simulation, the last two treatments are relaxed.
The color charge density possesses a finite thickness and evolves dynamically according to the continuity equation. 
In this model, which accounts for color charge densities with finite thickness, there is scope to include longitudinal spatial fluctuations within the color charge density. 
The longitudinal spatial fluctuations are taken into account through the longitudinal correlation length in the classical color charge density, a parameter in the model. 
Although the physical origin of this correlation is still non-trivial at this point, 
it is interesting to conduct simulations with various correlation lengths and numerically investigate their impact on the evolution of pressure.

The research is significant as it delves into two critical areas: 
firstly, it is vital for comprehending how closely the glasma reaches a hydrodynamic fluid within the classical approximation, especially in relation to its impact on the evolution of the degree of the pressure isotropy. 
{While pressure isotropy is a reliable indicator for tracking the thermalization of the system, 
it is crucial to acknowledge that, even when full pressure isotropy is achieved, the classical field is still far from quantum equilibrium.
Classical field theory does not reach quantum thermal distribution but exhibits the Rayleigh--Jeans divergence in its equilibrium state.
Therefore, it is not a suitable approximation for a quantum system nearing a thermal state, such as a hydrodynamic fluid.}
Secondly, it provides essential insights for developing more realistic models of the initial stages of the collision.

This article is organized as follows: 
In Section~\ref{Sec:Method}, we introduce the 3+1D glasma simulation method that we have recently developed~\cite{3Dglasma6}. 
In Section~\ref{Sec:Result}, we present the results of both transverse and longitudinal pressure during the evolution of the 3D glasma, comparing pressure isotropization across various longitudinal correlation lengths inside the colliding nuclei. 
Section~\ref{Sec:Summary} is devoted to providing a summary and perspectives on \mbox{our findings.}

\section{Method}\label{Sec:Method}
In what follows, we shall explain the description of the 3D glasma that we have recently developed~\cite{3Dglasma6}. 
We first provide a brief review of the CGC effective theory for a single relativistic nucleus. 
Subsequently, we discuss the understanding of the boost-invariant glasma created in the collision of relativistic nuclei using the eikonal approximation. 
Then, we introduce the 3+1D glasma simulation method, in which several aspects of the eikonal approximation are relaxed.

\subsection{Cgc Effective Thoery for Single Relativistic Nucleus}
In the CGC effective theory, effective for large nuclei at relativistic speeds, we treat `hard' partons, gluons and quarks with higher momentum than soft gluons, and soft gluons separately. 
Hard partons are modeled as classical color charge density $\rho$, while soft gluons are treated as the CYM field $A_\mu$ radiated from the classical color current $J_\mu$, determined by solving the CYM equations of motion with $J$, $[D_\mu,F^{\mu\nu}]=J^\nu$.
A covariant derivative is defined as $D_\mu \equiv \partial_\mu-igA_\mu$, and the field strength is defined by the commutator of $D$, $F_{\mu\nu}\equiv i[D_\mu,D_\nu]$.
In this context, only the dominant component of the classical color current in the infinite momentum limit is considered, $J^\mu=\delta^{\mu+} \rho/g$, where light-cone coordinates are introduced as $x^\mp = (t \mp z)/\sqrt{2}$.
The dominance of the $+$ component in the classical current is evident because the boost transformation in the positive $z$ direction enhance this component, $J=(J^+,J^-,\bm{J}_\perp) \to ( e^{\phi} J^+, e^{-\phi} J^-,\bm{J}_\perp)$, where $\phi$ is a rapidity.
Now the continuity equation for the classical color current can be simplified, $[D_\mu,J^\mu]=0 \to [D_+, \rho]=[\partial_+-igA_+,\rho]=0$.
Under the gauge condition of $A_+=A^-=0$, the continuity equation becomes $\partial_+ \rho=0$, and thus $\rho$ should be static in the light-cone time $x^+$,
\begin{align}
\rho=\rho(x^-,\bxp)\ .\label{Eq:rho_single}
\end{align}

The 
 soft gauge field, a solution of the equation of motion, $[D_\mu,F^{\mu\nu}]=J^\nu$, is a functional of $\rho$ and is known as
\begin{align}
A_\pm = 0\ ,\ \ A_i = \frac{i}{g} V \partial_i V^\dagger\ ,\label{Eq:A_single}
\end{align}
where the index $i$ runs over the transverse components, $i=1,2$, and $V$ is the Wilson line,
\begin{align}
V^\dagger(x^-,\bxp) = P_{x^-} \exp{\left[ -i \int^{x^-}_{-\infty} dx'^- \partial^{-2}_\perp \rho_{\rm cov}(x'^-,\bxp) \right]}\ .\label{Eq:V}
\end{align}

Here, $\rho_{\rm cov}$ denotes the classical color charge density in the covariant gauge condition and is related to the color charge density in the $A_- = 0$ gauge condition through the gauge transformation,
\begin{align}
\rho_{\rm cov} = V^\dagger \rho_{_{\rm (A_-=0)}} V\ .\label{Eq:gt}
\end{align}

The form of the classical color charge density in the CGC is not uniquely determined; instead, the weight function of the classical color charge density, $W_Y[\rho]$, is considered. 
The expectation value of a given observable that the soft gauge field possesses can be examined as its average under the weight function, 
\begin{align}
\average{\mathcal{O}} = \int \mathcal{D}\rho W_Y[\rho] \mathcal{O}[\rho]\ ,\label{Eq:eveave}
\end{align}
which depends on the separation scale $Y$ between hard partons and soft gluons. 
When the collision energy is not too high, the classical color charge density consists only of valence quarks. 
In the ideal situation, where a nucleus is extremely dense and extended infinitely in the transverse direction perpendicular to the collision axis, 
the weight function for such a color charge density consisting of valence quarks is provided as a Gaussian function~\cite{MV1,MV2,MV3,Kov}, known as the McLerran--Venugopalan (MV) model,
\begin{align}
W_Y[\rho(x^-,\bxp)] \propto \exp{\left\{-\frac{{\rm Tr}[\rho(x^-,\bxp)^2]}{2[g^2\mu(x^-,\bxp)]^2}\right\}}\ ,\label{Eq:MV}
\end{align}
where $(g^2\mu)^2$ is the averaged square of the color charge density, which is a parameter controlling the magnitude of $\rho$ in the MV model, and its integration, $\int d\xm (g^2\mu)^2$, should be taken in the same order of the squared saturation scale, $\qs^2$, characterizing the CGC.
It should be noted that the weight function in the MV model does not depend on the gauge choice, which is ensured by the trace in Equation~\refeq{Eq:MV}.
The expectation value of the single color charge density in the MV model vanishes, 
\begin{align}
\average{\rho^a(x^-,\bxp)}=0\ ,
\end{align}
where the index $a$ represents the internal degrees of freedom of $SU(N)$, 
and the expectation value of two color charge densities in the MV model satisfies a relation, suggesting $\rho$ at different spatial points are independent of each other,
\begin{equation}
\begin{split}
&\average{ \rho^a(x^-,\bxp)\rho^b(x'^-,\bxp') }\\
&= \delta^{a,b} \left( g^2\mu(x^-,\bxp) \right)^2 \delta(x^--x'^-) \delta^2(\bxp-\bxp')\ .\label{Eq:mv_rho}
\end{split}
\end{equation}

As the collision energy increases, $\rho$ should incorporate contributions from an increasing number of particles with large momentum, which can be captured by solving the JIMWLK equation~\cite{JIMWLK1,JIMWLK2,JIMWLK3,JIMWLK4,JIMWLK5,JIMWLK6,JIMWLK7,JIMWLK8}, the evolution equation for the separation scale, $Y$, with an initial condition based on the MV model.
In the actual application for studying observables in the CGC, we randomly generate an event-by-event color charge density according to the weight function, $W_Y$, and estimate the expectation values of observables as event average values.

Later, we consider the color charge density based on the MV model, which is relevant for the energy of Au--Au collisions at RHIC, with a center-of-mass energy per nucleon pair of 200 GeV. 
Effects beyond the MV model, which can be taken into account by solving the JIMWLK equation, become more important when analyzing larger energy regions at LHC.

\subsection{$3$D Glasma in Collision of Relativistic Nuclei}
We can describe the collision of two relativistic nuclei based on the CGC picture.
In the case of two nuclei, the classical color current is described as the sum of the two classical color charge densities, each representing different nuclei,
\begin{align}
J^{\mu} = \frac{1}{g} \delta^{\mu+} \rho^{(1)}
        + \frac{1}{g} \delta^{\mu-} \rho^{(2)}\ .\label{Eq:J_double}
\end{align}

The color charge density $\rho^{(1)}$ represents a nucleus moving in the positive $z$ direction, while the color charge density $\rho^{(2)}$ represents a nucleus moving in the negative $z$ direction. 

Let us show the solution of the CYM equation of motion and the continuity equation.
Before the two nuclei make contact, due to causality, a solution for the gauge field is simply given as the sum of the solutions for each single nucleus, 
\begin{align}
A_\pm = 0\ ,\ \ A_i = A^{(1)}_i + A^{(2)}_i
\end{align}
where $A^{(1/2)}_i$ is the gauge field radiated from $\rho^{(1/2)}$, 
\begin{align}
A^{(1/2)}_i              &= \frac{i}{g} V^{(1/2)} \partial_i V^{(1/2)\dagger}\ ,\\
V^{(1)\dagger}(x^-,\bxp) &= P_{x^-} \exp{\left[ -i \int^{x^-}_{-\infty} dx'^- \partial^{-2}_\perp \rho^{(1)}_{\rm cov}(x'^-,\bxp) \right]}\ ,\\
V^{(2)\dagger}(x^+,\bxp) &= P_{x^+} \exp{\left[ -i \int^{x^+}_{-\infty} dx'^+ \partial^{-2}_\perp \rho^{(2)}_{\rm cov}(x'^+,\bxp) \right]}\ .
\end{align}

However, the problem becomes much more complicated from the beginning time of the collision. 
After the collision happens, the CYM field includes the glasma emerging as a result of the collision.
Amazingly, the solution representing the glasma is known under the eikonal approximation. 
This approximation includes three treatments of the classical color charge density or current:
firstly, we keep the treatment considering only the dominant components of the classical color current, as in the single nucleus case, which is already assumed in Equation~\refeq{Eq:J_double};
secondly, we treat the classical color charge density as infinitely thin along the collision axis due to Lorentz contraction, $\rho^{(1/2)} \propto \delta(x^\mp)$;
and thirdly, we assume the classical color charge density is unchanged by the collisions, which means \mbox{two classical} color charge densities are static over the entire time, $\partial_\pm \rho^{(1/2)} = 0$.
Under the eikonal approximation, a regular solution of the CYM equation of motion, $[D_\mu,F^{\mu\nu}]=J^\mu$, at the infinitely small proper time, $\tau=0+$, is explicitly known as~\cite{glasma}
\begin{align}
A_i&= \lim_{x^- \to \infty} A^{(1)}_i + \lim_{x^+ \to \infty} A^{(2)}_i\ ,\ \ A_\eta=0\ ,\label{Eq:2Dsolution_A}\\
E^i&= 0\ ,\ \ E^\eta=\lim_{x^\mp \to \infty} ig[A^{(1)}_i, A^{(2)}_i]\ ,\label{Eq:2Dsolution_E}
\end{align}
where the Fock--Schwinger gauge condition $A^\tau=0$ is imposed, and the electric fields in this gauge are defined as ${E^i \equiv \tau \partial_\tau A_i}$ and ${E^\eta \equiv \partial_\tau A_\eta /\tau}$.
Here, Milne coordinates are introduced as $\left(\tau, \eta\right)=\left(\sqrt{2\xm\xp},\frac{1}{2}\ln{\frac{\xp}{\xm}}  \right)$, which are natural coordinates for describing the matter created in relativistic heavy-ion collisions. 
The relativistic nucleus exists around $\tau=0$, and the evolution of the created matter expanding along the collision axis at close to the speed of light can be naturally followed in terms of proper time, $\tau$, instead of \mbox{ordinary time, $t$.}

Some important points about the solution for $\tau=0+$ under the eikonal approximation, shown in Equations~\refeq{Eq:2Dsolution_A} and \refeq{Eq:2Dsolution_E}, should be mentioned. 
First, this solution represents only the glasma, since the gauge fields representing the soft gluon, the Weizs\"{a}cker--Williams field, are located solely at $\tau=0$ in the limit of the infinite thin nucleus. 
The subsequent evolution of the glasma at a larger $\tau$ can be tracked by solving the CYM equation of motion, using the $\tau=0+$ solution as the initial condition. 
Second, this solution exhibits exact boost invariance, independent of rapidity, $\partial_\eta A=0$ and $\partial_\eta E=0$. 
Third, the color electromagnetic fields in this boost-invariant glasma orient along the collision axis, $E^\eta,B^\eta \neq 0$ and $E^i,B^i = 0$.
These anisotropic color electromagnetic fields lead to the longitudinal pressure and the transverse pressure, defined as $P_{\rm \perp} \equiv \tau^2 T^{\eta\eta}$ and $P_\perp \equiv (T^{11}+T^{22})/2$, having the same absolute value but opposite signs, $P_{\rm \perp}=-P_{\rm L}$.

\textls[-15]{Recently, much attention has been paid to the 3+1D glasma simulation, which goes beyond the eikonal approximation. 
In the simulation methods proposed in previous \mbox{papers~\cite{3Dglasma1,3Dglasma2,3Dglasma3,3Dglasma4,3Dglasma5,3Dglasma6},}} the eikonal approximation is relaxed, allowing the classical color charge density to possess a finite thickness and evolve dynamically according to the continuity equation. 
Among these methods, we employ the 3+1D glasma simulation method that we have recently developed~\cite{3Dglasma6}.
The distinctive feature of our 3+1D glasma simulation method is its use of Milne coordinates to track the evolution of the $3$D glasma, whereas other simulation methods employ Minkowski coordinates. 
In what follows, we review our 3+1D glasma simulation method in Milne coordinates.

In the 3+1D glasma model, the color charge density is no longer assumed to be infinitely thin but has a finite thickness in the direction of the collision axis. 
Additionally, the color charge density dynamically evolves according to the continuity equation for each nucleus. 
Since it is not practical to solve a set of CYM equations of motion and continuity equations analytically without any assumptions, we numerically solve these equations in Milne coordinates with the initial collision conditions set for the proper time before the collision occurs. 
It is important to note that, in this method, we set the center positions of each nucleus in light-cone coordinates not around $\xmp=0$ but around positive finite values, $\xm=x^{(1)}_{\rm c}>0$ and $\xp=x^{(2)}_{\rm c}>0$.
The model parameters $x^{(1)}_{\rm c}$ and $x^{(2)}_{\rm c}$ can be any value as long as they are sufficiently large. 
Consequently, the two nuclei do not yet make contact at the proper time of $0+$.
To distinguish these Milne coordinates from the usual ones, which are defined so that the centers of mass for each nucleus coincide at $\tau=0$, we introduce new notation, $\left(\ttau, \teta\right)=\left(\sqrt{2\xm\xp},\frac{1}{2}\ln{\frac{\xp}{\xm}}  \right)$.
Accordingly, the definition of the usual Milne coordinates is now changed to $\left(\tau, \eta\right)=\left(\sqrt{2(\xm-x^{(1)}_{\rm c})(\xp-x^{(2)}_{\rm c})},\frac{1}{2} \ln{\frac{\xp-x^{(2)}_{\rm c}}{\xm-x^{(1)}_{\rm c}}}  \right)$.

The initial condition of the modified proper time used in the 3+1D glasma simulation, $\ttau_{\rm ini}$, is set to be significantly earlier than the moment when the centers of mass for each nucleus coincide. 
Causality implies that the initial condition of the classical color charge density and CYM field at $\ttau=\ttau_{\rm ini}$ is given by the sum of the solutions of the evolution equations for each individual nucleus.
As shown in Equation~\refeq{Eq:rho_single}, the classical color charge density before the collision is static in the light-cone time.
Therefore, the initial classical color charge densities are the same as those in the infinite past,
\begin{align}
\lim_{x^{\pm} \to -\infty} \rho^{(1/2)}(x) = \rho^{(1/2)}_{\rm ini}(\xmp,\bxp)\ .
\end{align}

The way of giving the classical color charge density in the infinite past is explained in the next section.
The initial CYM field is given by the sum of the solutions of the CYM equation of motion for each individual nucleus, as shown in Equation~\refeq{Eq:A_single},
\begin{align}
A_i       &= A^{(1)}_i    + A^{(2)}_i \ ,\label{Eq:CYMini0}\\
A_{\teta} &= \xp A^- - \xm A^+ = 0\ ,\\
A_{\ttau} &= \frac{\xp}{\ttau} A^- + \frac{\xm}{\ttau} A^+ = 0\ ,
\end{align}
where $(A_{\ttau},A_{\teta})$ are related to $(A^-,A^+)$ through the general coordinate transformation.
This solution satisfies the Fock--Schwinger gauge condition for the modified proper time, $A_{\ttau}=0$.
Accordingly, the electric fields in this modified Fock--Schwinger gauge condition are given by
\begin{align}
E^i       &= \ttau \partial_{\ttau} A_i
           =       x^- \partial_- A^{(1)}_i + x^+\partial_+ A^{(2)}_i\ ,\label{Eq:CYMini1}\\
E^{\teta} &=       \frac{1}{\ttau} \partial_{\ttau} A_{\teta} =0\ .
\end{align}

However, in the actual setup for calculations performed later, we assume that the color charge densities before the collision have a Gaussian shape. 
Since the Gaussian function extends to infinity, two nuclei are correlated with each other, even in the infinite past.
Therefore, the simple sum of the solutions for each single nucleus, as described above, is no longer a solution of the evolution equations and violates Gauss's law,
\begin{align}
\sum_{j=1,2,\teta} [D_j, E^j]=J^{\ttau}=\frac{1}{\ttau}\left[ \xm J^+ + \xp J^- \right]\ .
\end{align}

In fact, we can make the electric fields given above comply with Gauss's law by adding the longitudinal electric field, the expression of which is the same as that for the boost-invariant~glasma,
\begin{align}
E^{\tilde{\eta}} = ig[A^{(1)}_i, A^{(2)}_i]\ .\label{Eq:CYMini2}
\end{align}

The additional longitudinal electric field is negligibly small, $E^{\tilde{\eta}} \ll E^i$, because the overlap of the tails of the Gaussian-shaped color charge densities is very minor at $\ttau=\ttau_{\rm ini}$.
Finally, the initial condition of the classical color charge density and CYM field we employ is given by the sum of the solutions of the evolution equations for each individual nucleus with the small modification for the longitudinall electric field,
\begin{align}
\rho^{(1/2)} &= \rho^{(1/2)}_{\rm ini}\ ,\\
A_i          &=                A^{(1)}_i +               A^{(2)}_i\ ,\ \ \ A_{\teta} = 0\ ,\\
E^i          &= x^- \partial_- A^{(1)}_i + x^+\partial_+ A^{(2)}_i\ ,\ \ \ E^{\teta} = ig[A^{(1)}_i, A^{(2)}_i]\ .
\end{align}

In actual calculations, we place the CYM field and classical color current on a spatial lattice and track their evolution using their respective evolution equations. 
The detailed formulation of these is provided in our previous paper~\cite{3Dglasma6}.

\subsection{3D Color Charge Density}\label{Sec:ColorCharge}
We assume the classical color charge density in the infinite past as the initial condition of the color charge density, $\rho_{\rm ini}$, under the premise that each nucleus is still largely unaffected by the other at the initial simulation time, $\ttau_{\rm ini}$. 
To simulate the effect of non-trivial longitudinal correlation in the classical color charge density, the initial condition, $\rho_{\rm ini}$, is modeled using the modified MV model. 
In this model, $\rho_{\rm ini}$ is given to satisfy the following relation, the generalization of Equation~\refeq{Eq:mv_rho} by replacing the delta function for $x^\mp$ with a Gaussian function,
\vspace{-12pt}\begin{adjustwidth}{-\extralength}{0cm}
\centering 
\begin{align}
\average{ \rho^{(1/2)a}_{\rm ini}(x^\mp,\bxp)\rho^{(1/2)b}_{\rm ini}(x'^\mp,\bxp') }
= \delta^{a,b} \left( g^2\mu^{(1/2)}(\frac{x^\mp+x'^\mp}{2},\bxp) \right)^2
        N(x^\mp - x'^\mp,l^{(1/2)}_{\rm L})
        \delta^2(\bxp-\bxp')\ ,\label{Eq:rho_ave1}
\end{align}
\end{adjustwidth}
where $N(\xmp,l^{(1/2)}_{\rm L})$ denotes the normal distribution with a variance of $l^{(1/2)}_{\rm L}$, {which represents the longitudinal shape of nucleus in the $x^\mp$ coordinate}, and $g^2 \mu^{(1/2)}$ becomes a function of the center position of the two color charge densities. 
The variances of the Gaussian functions, $l^{(1/2)}_{\rm L}$, act as longitudinal correlation lengths. 
For simplicity, we further assume that the averaged square of the color charge density also has a Gaussian shape, 
\begin{align}
\left( g^2\mu^{(1/2)}(\frac{x^\mp+x'^\mp}{2},\bxp) \right)^2 = \left( g^2\bar{\mu}^{(1/2)}(\bxp) \right)^2 N(\frac{x^\mp+x'^\mp}{2}-x^{(1/2)}_{\rm c},r^{(1/2)}_{\rm L})\ ,\label{Eq:rho_ave2}
\end{align}
where $r^{(1/2)}_{\rm L}$ are parameters that act as longitudinal extensions of a nucleus in the light-cone coordinates.
In the actual simulations, we consider the collision of the same kinds of a nucleus, $g^2\bar{\mu}^{(1)}=g^2\bar{\mu}^{(2)}=g^2\bar{\mu}$, $l^{(1)}_{\rm L}=l^{(2)}_{\rm L}=l_{\rm L}$ and $r^{(1)}_{\rm L}=r^{(2)}_{\rm L}=r_{\rm L}$, and take $\xc^{(1)}=\xc^{(2)}=\xc$ for simplicity.

Under the setup shown in Equation~\refeq{Eq:rho_ave2}, the event-by-event color charge density satisfying the relation shown in Equation~\refeq{Eq:rho_ave1} as an event average can be obtained by using the event-by-event random number $\Gamma$,  
\begin{align}
\rho^{(1/2)}_{\rm ini}(x^\mp,\bxp) = N(x^\mp-x_{\rm c}, \sqrt{2} r_{\rm L}) \Gamma^{(1/2)}(x^\mp,\bxp)\ ,
\end{align}

The random number $\Gamma^{(1/2)}$ satisfies the following event average,
\vspace{-12pt}\begin{adjustwidth}{-\extralength}{0cm}
\centering 
\begin{align}
\average{ \Gamma^{(1/2)a}_i(x^\mp,\bxp) \Gamma^{(1/2)b}_i(x'^\mp,\bxp') }
= 
\delta^{a,b}
\sqrt{2\pi\left(2r^2_{\rm L}+\sigma^2_{\rm L}\right)}
\left( g^2\bar{\mu} \right)^2
N(x^\mp-x'^\mp,\sigma_{\rm L}) \delta^2(\bxp-\bxp')\ ,\label{Eq:gamma_i}
\end{align}
\end{adjustwidth}
with
\begin{align}
\sigma^{-2}_{\rm L} = l^{-2}_{\rm L} - (2 r_{\rm L})^{-2}\ .
\end{align}

To make $\sigma_{\rm L}$ positive, $l_{\rm L}$ should be smaller than $2 r_{\rm L}$.
This upper bound of the correlation length seems to be consistent with the spatial expansion of the color charge density.
The detailed derivation of the relation between $\Gamma$ and $\rho_{\rm ini}$ is given in our previous paper~\cite{3Dglasma6}.

To avoid the infrared singularity of the Wilson line stemming from $\partial^{-2}_\perp \rho$, we introduce an infrared cutoff, $\Lambda_{\rm IR}$, by adding the following factor in front of the Fourier transform of the classical color charge density,
\begin{align}
\rho_{\rm ini}(\bk) \to  \frac{k^2_\perp}{k^2_\perp + \Lambda^2_{\rm IR}} \rho_{\rm ini}(\bk)\ ,
\end{align}
where $k_\perp$ is the transverse momentum.
The ultraviolet cutoff, $\Lambda_{\rm UV}$, is also introduced. 
{The UV cutoff suppresses the high transverse momentum modes of the color charge density that exceed the limit of the CGC description. 
The color charge density represents the aggregate of color charges from hard partons within a transverse resolution scale, 
which is approximately the inverse of the transverse momentum. 
Therefore, UV modes originating from color charges in a small transverse spatial region are not so dense that the classical treatment is justifiable.
Furthermore, the UV cutoff plays a role in facilitating the taking of the limit in continuous space.
It is shown by an analytic calculation using the MV model that, without UV cutoff, a local operators of two gauge fields diverge in continuous space~\cite{CFKL}.}
The ultraviolet cutoff should be chosen as a larger value than the saturation scale, $\qs$, which is the scale focused on in the CGC.
{A detailed analysis of the UV cutoff dependence is beyond the scope of this paper.}

\section{Numerical Results}\label{Sec:Result}
We show the numerical results for the 3D glasma, computed using the modified Milne coordinates. 
Our simulations are conducted using a lattice framework {with a link variable, 
which is a gauge covariant and ensures that Gauss's law holds under their evolution equations~\cite{3Dglasma6}.}
We utilize the leap-frog method to solve {those} equations set on the lattice. 
As for the boundary conditions, we employ periodic boundaries in the transverse directions, 
while ensuring that both the CYM field and classical color current are set to zero at the longitudinal direction's boundaries.
{The number of colors is $2$, not $3$.
While calculations based on SU($2$) cannot be used to make quantitative predictions for relativistic heavy-ion experiments, 
they are expected to lead to results qualitatively similar to SU($3$).}

In what follows, we first present the set of parameters involved in the model. 
Then, we provide the numerical results for the pressure and energy density at mid-rapidity, as well as the pressure results in the broader rapidity region. 
The energy density, transverse pressure, and longitudinal pressure are defined by the energy--momentum tensor in the 'usual' Milne coordinates,
\begin{align}
\varepsilon &\equiv \frac{\int d^2x_\perp \average{T^{\tau\tau} }}{\int d^2x_\perp}\ ,\\
P_\perp     &\equiv \frac{\int d^2x_\perp \average{\frac{1}{2}\left(T^{11}+T^{22}\right)}}{\int d^2x_\perp}\ ,\\
P_{\rm L}   &\equiv \frac{\int d^2x_\perp \average{\tau^2 T^{\eta\eta}}}{\int d^2x_\perp}\ .
\end{align}

Here, they are integrated values over the transverse plane since we focus on their dependence on rapidity.
Reflecting the fact that the energy--momentum tensor of the CYM field is traceless due to its conformal symmetry, the relationship between energy and pressure, $\varepsilon=2P_\perp+P_{\rm L}$, is held.
{Since the actual simulation is performed in modified Milne coordinates, the energy--momentum tensor to be calculated directly is in modified Milne coordinates $T^{\ttau\ttau}, T^{\ttau\teta}$ and $T^{\teta\teta}$.
Therefore, the energy--momentum tensor in usual Milne coordinates $T^{\tau\tau}$ and $T^{\eta\eta}$ can be obtained by a general coordinate transformation,}
\begin{equation}
\begin{split}
T^{\tau\tau} 
&= \frac{1}{\ttau^2-2\ttau \bar{x}_{\rm c}\cosh{\teta}+\bar{x}^2_{\rm c}}\Big[
 \left(\ttau-\bar{x}_{\rm c}\cosh{\teta}\right)^2 T^{\ttau\ttau}\\
&+\left(\bar{x}_{\rm c}\sinh{\teta}\right)^2 \ttau^2 T^{\teta\teta}
 -\bar{x}_{\rm c}\sinh{\teta}\left(\ttau-\bar{x}_{\rm c}\cosh{\teta}\right)\ttau T^{\ttau\teta}\Big]\ ,
\tau^2 T^{\eta\eta}
\end{split}
\end{equation}
\vspace{-12pt}\begin{equation}
\begin{split}
&= \frac{1}{\ttau^2-2\ttau \bar{x}_{\rm c}\cosh{\teta}+\bar{x}^2_{\rm c}}\Big[
 \left(\bar{x}_{\rm c}\sinh{\teta}\right)^2 T^{\ttau\ttau} \\
&+\left(\ttau-\bar{x}_{\rm c}\cosh{\teta}\right)^2 \ttau^2 T^{\teta\teta}
 -\bar{x}_{\rm c}\sinh{\teta}\left(\ttau-\bar{x}_{\rm c}\cosh{\teta}\right)\ttau T^{\ttau\teta} \Big]\ ,
\end{split}
\end{equation}
where $\bar{x}_{\rm c}=\sqrt{2} \xc$.
{To estimate the energy--momentum tensor from lattice simulations, we use the definition of the energy--momentum tensor on a lattice given in our previous paper~\cite{3Dglasma6}.}

\subsection{Model Parameters}

\begin{figure}[H]
\includegraphics[width=6cm]{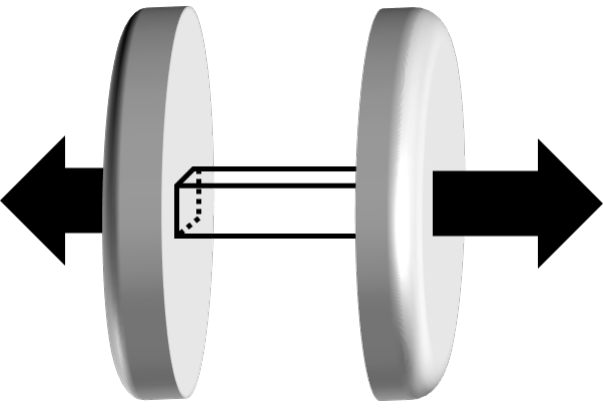}
\caption{
Our setup simulates the system in the central region of nuclei in the transverse plane.
\label{Fig:center}
}
\end{figure}

The set of model parameters for the initial $3$D color charge density includes the following: 
the longitudinal extension of the classical color charge density, $r_L$, 
the longitudinal correlation length in the classical color charge density, $l_L$, 
the parameter controlling the averaged square of the color charge density, $g^2\bar{\mu}$, 
and the infrared and ultraviolet cutoffs, $\Lambda_{\rm IR}$ and $\Lambda_{\rm UV}$. 
To simulate the Au--Au collision with a center-of-mass energy per nucleon pair of $200$ GeV at RHIC, 
we adjust $\sqrt{2} r_L$ to roughly match the radius of a gold nucleus divided by the gamma factor $\gamma=108$ and the factor $\sqrt{2}$ stemming from the definition of the light-cone variables, $r_L=0.1/\qs$, 
where the saturation scale is set at $\qs=1$ GeV. 
The parameter $g^2\bar{\mu}$ is set to ensure that the integration of the averaged square of the color charge density over the light-cone variable coincides with the squared saturation scale, $\int z \left( g^2\mu(x,\bxp) \right)^2=Q^2_s$.
This setup for $g^2\bar{\mu}$ indicates that the event average of a given observable has a translational invariance in the transverse plane. 
Such a translational invariance is approximately established in the central region of large nuclei in the transverse plane, where the thickness of the nucleus does not vary largely. 
Therefore, our setup simulates the glasma in the central region, as depicted in Figure~\ref{Fig:center}. 
The infrared cutoff is set at $\Lambda_{\rm IR}=0.2 \qs$, and the ultraviolet cutoff is taken to be double the value of $\qs$, $\Lambda_{\rm UV}=2\qs$. 
It is important to note that in the classical approximation the choice of the coupling constant only provides the trivial change in the field variables, as $A,E,J \propto 1/g$.
This is because the coupling constant can be excluded in both the classical equations of motion and the continuity equation by redefining the scaled fields, $A'=gA, E'=gE$, and $J'=gJ$.
The numerical results shown in the later sections are normalized by $g$ to make them independent of $g$.

\vspace{-3pt}  

We explore the effect of longitudinal correlation by comparing the 3+1D glasma under various correlation lengths, $l_{\rm L}$, ranging from $0.01r_{\rm L}$ to $2r_{\rm L}$. 
The smallest value, $l_{\rm L}=0.01r_{\rm L}$, which is significantly less than the nucleus radius, is thought to arise from the intricate structure within a nucleon. 
The largest value, $l_{\rm L}=2r_{\rm L}$, which is twice the nucleus radius, is the maximum limit discussed in Section~\ref{Sec:ColorCharge}. 
It is crucial to emphasize that the physical origin of the longitudinal correlation is beyond the scope of this paper.

In addition, the following parameters should be inputted for the 3D glasma simulation.
These parameters are chosen to be sufficiently safe values to minimize their impact on the physical results.
The transverse lattice grid number and lattice spacing are set to $N_\perp=56$ and $a_\perp=0.4/Q_s$, respectively. 
For the lattice in the modified rapidity direction $\teta$, the grid number and spacing are $L_{\teta}=896, 7168$ and $a_{\teta}=2.6/L_{\teta}$, respectively. 
It is noteworthy that simulations with smaller $l_{\rm L}$ require smaller longitudinal lattice spacings to accurately capture the structure of the color charge density in the longitudinal direction. 
The beginning modified proper time in the simulation and the center positions of the nuclei are taken as $\ttau_{\rm ini}=32 r_{\rm L}$ and $\xc = 20\sqrt{2}r_{\rm L}$, respectively. 
All results in this section are estimated using $48$ events. 
The error in the results is estimated by dividing the unbiased variance by the square root of the number of events. 
According to the central limit theorem, the unbiased variance aligns with the variance of the mean value in the infinitely large number of events.

\subsection{Midrapidity Region}

\begin{figure}[H]
\includegraphics[width=10cm]{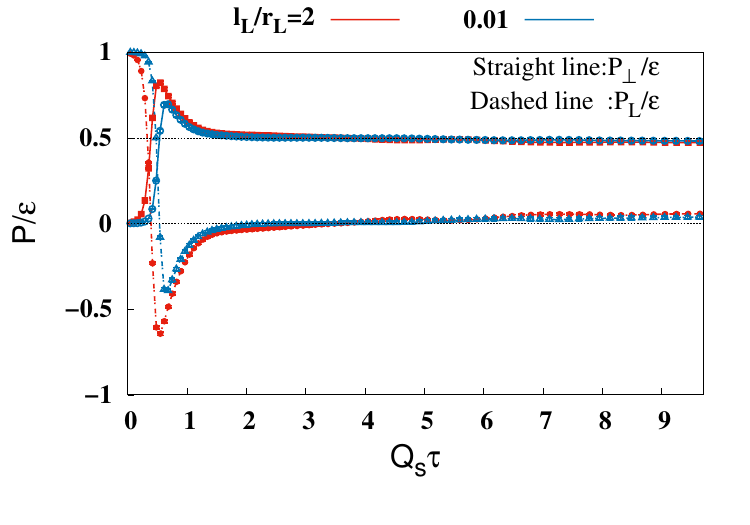}
\caption{
Evolution of the transverse and longitudinal pressure normalized by the energy density, $P_\perp/\varepsilon$ and $P_{\rm L}/\varepsilon$, in the mid-rapidity, $\eta=0$.
The red and blue lines show the results for $l_{\rm L}/r_{\rm L}=2$ and $0.01$, respectively.
The proper time is normalized by the saturation scale as $\qs \tau$.
\label{Fig:Pressure_mid}
}
\end{figure}

First, we investigate the behavior of pressure and energy density in the mid-rapidity region, $\eta=0$. 
Figure \ref{Fig:Pressure_mid} illustrates the evolution of glasma pressure at $\eta=0$, with the pressure normalized to the energy density.
The time range $0 \leq \qs \tau \leq 9.7$ covers the time before the onset time of the hydrodynamics, $\tau=0.6 \sim 1.0$~\cite{Heinz}, as predicted by the experimental analysis.
The blue and red lines indicate pressures resulting from collisions of nuclei with two different lengths of longitudinal correlation in color charge densities, $l_{\rm L}=0.01r_{\rm L}$ and $2r_{\rm L}$.
The solid and dotted lines correspond to transverse and longitudinal pressures over the energy density, $P_\perp/\varepsilon$ and $P_{\rm L}/\varepsilon$, respectively.

\vspace{-3pt}

Despite exploring two distinct longitudinal correlations in the color charge densities, the observed pressures were qualitatively similar, though they showed quantitative differences, particularly in the early time region.
Notably, at $\qs \tau=0$, transverse pressure is minimal compared to longitudinal pressure, reflecting the overlap of two nuclei centers at $(\tau,\eta)=(0,0)$ and the dominance of the Weizs\"{a}cker--Williams field, a transverse electromagnetic field representing nuclei, which significantly contributes to longitudinal but not transverse pressure.
Between $\qs \tau=0$ and $0.6$, both transverse and longitudinal pressures undergo substantial changes, eventually acquiring opposite signs.
The realization of the relation $P_\perp>0>P_{\rm L}$ indicates the formation of a strong longitudinal color electromagnetic field between separating nuclei, like the boost invariant glasma. 
From $\qs \tau=0.6$ to $2$, the transverse pressure trends towards $0.5$, while longitudinal pressure approaches zero, with no significant changes observed beyond $\tau=2$. 
These findings are in qualitative agreement with the previous results from the glasma calculations, where other effects of boost invariance breaking are taken into account~\cite{Instability14,Instability15,Instability16,Instability17,Instability18,Instability19,Instability20,Instability21,3DJIMWLK}.

\begin{figure}[H]
\includegraphics[width=10cm]{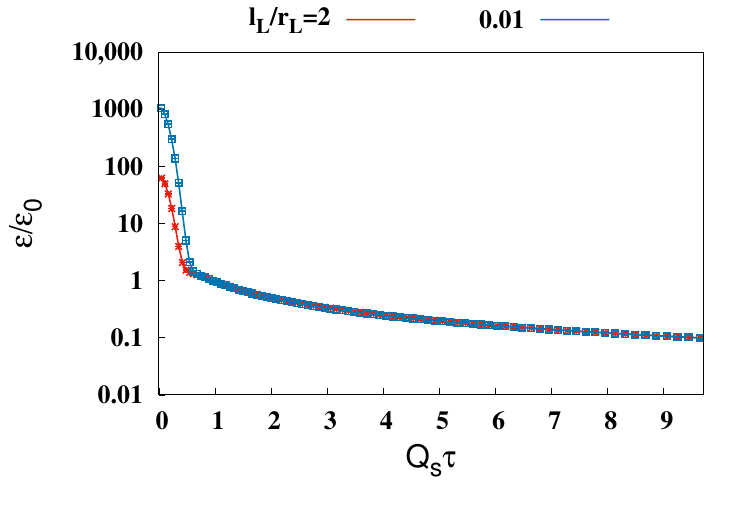}
\caption{
Evolution of the energy density $\varepsilon$ in the mid-rapidity, $\eta=0$, normalized by that for $\qs \tau=1$, $\varepsilon_0$.
The red and blue lines show the results for $l_{\rm L}/r_{\rm L}=2$ and $0.01$, respectively.
The proper time is normalized by the saturation scale as $\qs \tau$.
\label{Fig:Energy_mid}
}
\end{figure}

Figure~\ref{Fig:Energy_mid} illustrates the evolution of the glasma energy density $\varepsilon$ at $\eta=0$, normalized by that for $\qs \tau=1$, $\varepsilon_0$, where the solid blue and red lines depict the energy density arising from collisions of nuclei with two different lengths of longitudinal correlation in color charge densities, $l_{\rm L}=0.01r_{\rm L}$ and $2r_{\rm L}$.
The time range is taken to be the same as in Figure~\ref{Fig:Pressure_mid}.
As in the pressure isotropization in Figure~\ref{Fig:Pressure_mid}, despite the choice of $l_{\rm L}$, the results of the $3$D glasma simulation are the same qualitatively at late time.
Between $\qs \tau=0$ and $0.6$, the energy density is dominated by the Weizs\"{a}cker--Williams field rather than the glasma. 
After $\qs \tau=0.6$, as the nuclei move past $\eta=0$ the evolution of the glasma's energy density becomes apparent. 
A power function fitting with fitting parameters $A$ and $B$, $f_{\rm fit}(\tau)=A/\tau^{B}$, was employed to accurately determine the decay rate of the glasma energy density in the later stages. 
The obtained fitting parameter $B$, $B=1.06 \pm 0.001$ for $l_{\rm L}=0.01r_{\rm L}$ and $B=1.04 \pm 0.0004$ for $l_{\rm L}=2r_{\rm L}$, indicates that the energy of the glasma at $\eta=0$ decreases at almost the inverse of the proper time, $1/\tau$.
This $1/\tau$ decrease in the energy density at late time has been observed by previous $2$D and $3$D glasma calculations~\cite{Instability14,Instability16,Instability22}.

The negligible longitudinal pressure, $P_\perp \gg P_{\rm L}$, and the decay of energy density, $\epsilon \propto 1/\tau$, in the late stage, as observed above, are consistent with the system expanding longitudinally like a dilute gas. 
These observations significantly deviate from the behavior expected in an ideal gas, where pressure is isotropic, $P_\perp = P_{\rm L}=P$, and the energy density decay is predicted by Bjorken hydrodynamics using the ideal gas equation of state, $\varepsilon=3P$, $\varepsilon \propto 1/\tau^{4/3}$.


\subsection{Broader Rapidity Region}

\begin{figure}[H]
\begin{adjustwidth}{-\extralength}{-4.5cm}
\hspace{-20pt}\includegraphics[width=6.8cm]{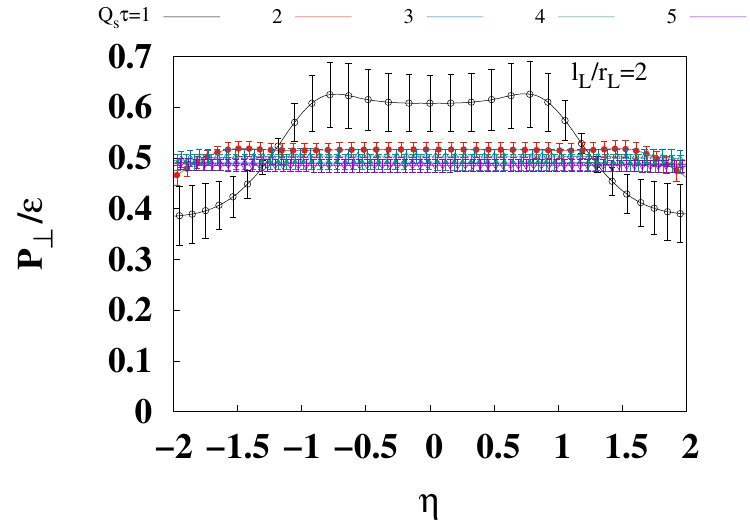}
\centering
\includegraphics[width=6.8cm]{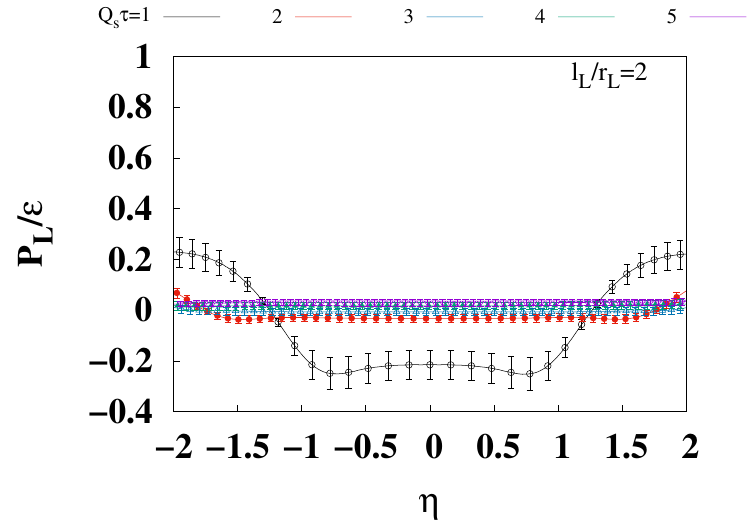}
\end{adjustwidth}
\begin{adjustwidth}{-\extralength}{-4.5cm}
\centering
\hspace{-20pt}\includegraphics[width=6.8cm]{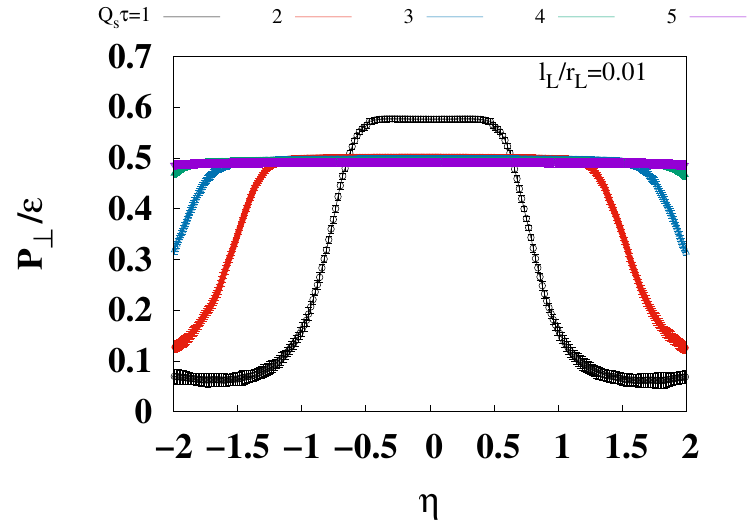}
\centering
\includegraphics[width=6.8cm]{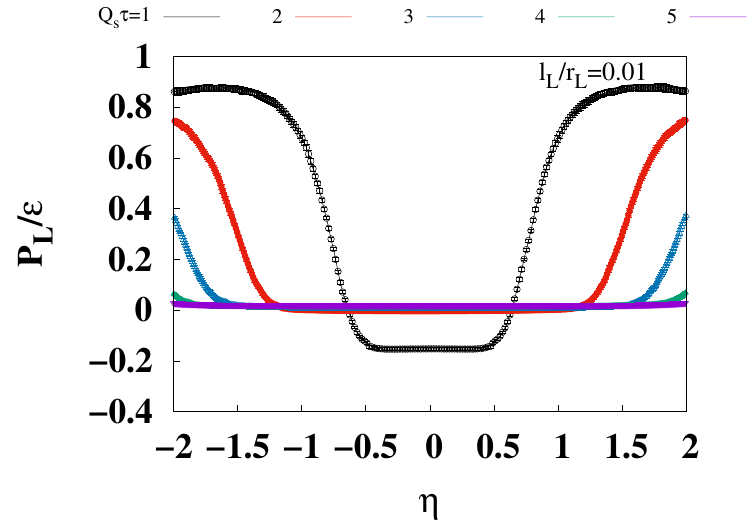}
\end{adjustwidth}
\caption{
The transverse and longitudinal pressures in the rapidity region, $-2 \leq \eta \leq 2$, for \linebreak $\qs=1,2,3,4$, and $5$.
The upper left and right panels show the transverse and longitudinal pressures for $l_{\rm L}/r_{\rm L}=2$, respectively.
The lower left and right panels show the transverse and longitudinal pressures for $l_{\rm L}/r_{\rm L}=0.01$, respectively.
The plots with different colors show the results for the different proper times, $\qs \tau=1,2,3,4$, and $5$.
\label{Fig:PressureTL}
}
\end{figure}

To deepen the understanding of the pressure isotropization in the $3$D glasma, we examine the pressure results across a broader rapidity region.
The Figure~\ref{Fig:PressureTL} provides both transverse and longitudinal pressures normalized by the energy density across the broader rapidity region, $-2\leq \eta \leq2$.
The upper left and right panels display the transverse and longitudinal pressures for $l_{\rm L}=2 r_{\rm L}$, respectively, 
and the lower panels show results for $l_{\rm }=0.01 r_{\rm L}$ in a similar arrangement.
The plots with different colors show the results for the different proper times, $\qs \tau=1,2,3,4$, and $5$.
It is found that around $\eta=0$ both transverse and longitudinal pressures over the energy density are relatively flat, undergoing drastic changes in a larger rapidity range.
To understand what contributes to the drastic change in behavior, we consider the evolution of the classical color current by examining its amplitude, defined as
\begin{align}
\rho_{\rm amp.} \equiv \frac{\int d^2x_\perp \average{{\rm Tr}\left[(\rho^{(1)})^2\right]+{\rm Tr}\left[(\rho^{(2)})^2\right] }}{\int d^2x_\perp}\ .\label{Eq:amp}
\end{align}

\begin{figure}[H]
\begin{adjustwidth}{-\extralength}{-4.5cm}
\includegraphics[width=6.8cm]{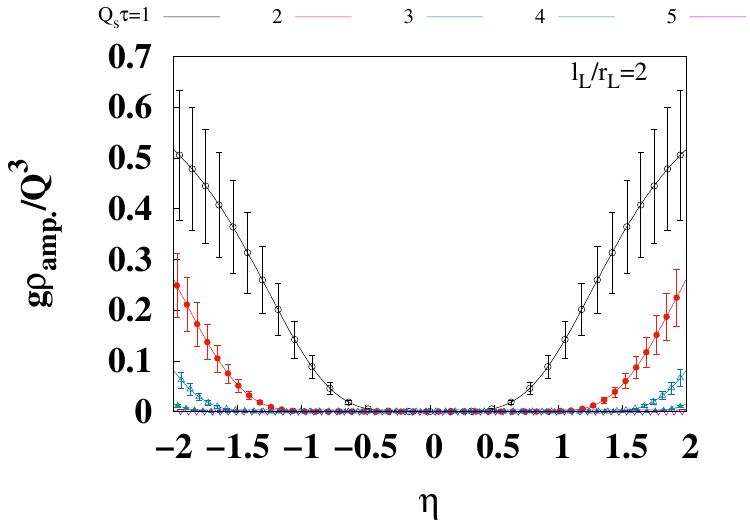}
\centering
\includegraphics[width=6.8cm]{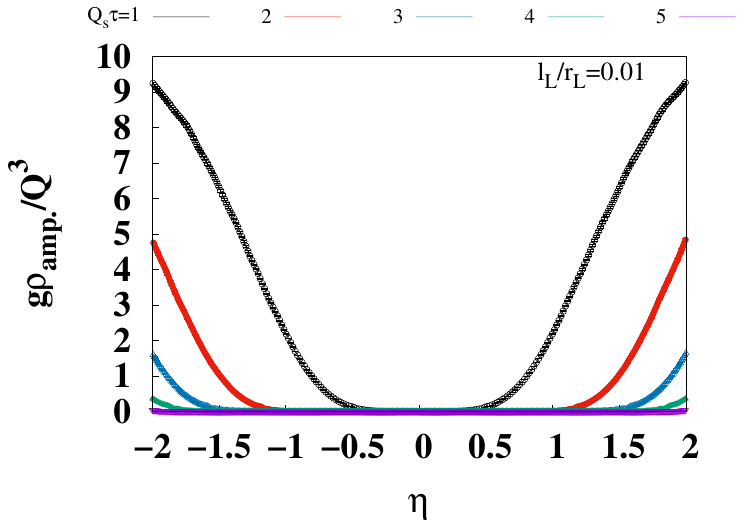}
\end{adjustwidth}
\caption{
The amplitude of the color charge density, defined in Equation~\refeq{Eq:amp}, in the rapidity region, $-2 \leq \eta \leq 2$, for $\qs=1,2,3,4$, and $5$.
The left and right panels show the results for $l_{\rm L}/r_{\rm L}=2$ and $0.01$, respectively.
The results are normalized by the saturation scale and coupling constant.
\label{Fig:J}
}
\end{figure}

It can be seen in Figure~\ref{Fig:J} that $\rho_{\rm amp.}$ vanishes in the rapidity region for flat $P_\perp/\epsilon$ and $P_{\rm L}/\epsilon$, shown in Figure~\ref{Fig:PressureTL}.
On the other hand, in the larger rapidity region, where $P_\perp/\epsilon$ and $P_{\rm L}/\epsilon$ change significantly, $\rho_{\rm amp.}$ is also found to increase significantly.
This correspondence indicates that the former and latter regions should be identified as the glasma and the mixture of the glasma and Weizs\"{a}cker--Williams field.
Therefore, the flat region in Figure~\ref{Fig:PressureTL} indicates that the pressure isotropization of the glasma at non-zero rapidity is almost the same as those \mbox{at mid-rapidity.}

\vspace{-8pt}

\section{Summary}\label{Sec:Summary}
We have investigated the effects of boost invariance breaking on the isotropization of pressure in the glasma. 
This breaking is caused by spatial fluctuations of the classical color charge density along the collision axis. 
Our calculation is performed using the 3+1D glasma simulation that we have recently developed.
In this method, the longitudinal spatial fluctuations are taken into account through the longitudinal correlation length in the color charge density, a model parameter.
While the physical origin of this correlation is still a subject of ongoing research, simulations with various correlation lengths offer useful insights into both the pressure evolution of the glasma and a deeper understanding of the $3$D glasma model itself.

We have provided numerical results for pressure and energy density at mid-rapidity and in a broader rapidity region. 
Our findings show that, despite variations in longitudinal correlation lengths, the evolutions of the pressure isotropy observed are qualitatively similar. 
At $\qs \tau=0$, the transverse pressure is minimal compared to longitudinal pressure, indicative of the dominance of the Weizs\"{a}cker--Williams field.
Then, at the early time, the positive transverse pressure and negative longitudinal pressure appear, which indicates the formation of strong longitudinal color electromagnetic fields between separating nuclei like the boost invariant glasma.
Finally, the transverse pressure becomes almost half of the energy density, and longitudinal pressure becomes negligible at the late time.
Further, it is found that, at this late time, the glasma's energy density at mid-rapidity decreases as $1/\tau$.
Both the negligible longitudinal pressure and the $1/\tau$ decay of energy density in the late stage are consistent with the system expanding longitudinally like a dilute gas,
deviating significantly from the Bjorken hydrodynamics with the ideal gas equation of state.
By examining the pressure results across a broader rapidity region, we further enhance the understanding of pressure isotropization. 
The pressure isotropization of the glasma in the non-zero rapidity region is almost the same as that observed in mid-rapidity.

Within the scope of this study, fluctuations originating from the correlation length of the color charge density do not facilitate the isotropization of pressure in the glasma. 
This suggests that a framework incorporating effects beyond the classical approximations, such as kinetic theory, is required to explain the hydrodynamization of the matter produced in heavy-ion collisions. 
However, note that our results do not imply that the setting of the correlation length is insignificant in the $3$D glasma model. 
In fact, the degree of pressure isotropization in the early stage and the magnitude of the normalized energy density are found to be dependent on the correlation length. 
Therefore, selecting an appropriate correlation length is likely essential for conducting more realistic $3$D glasma simulations.
In addition, as a potential area for future research, it would be intriguing to explore how the longitudinal correlation length of the color charge density is translated into rapidity correlations in experimental observables, such as elliptic flow.

We discuss the interpretation of the longitudinal correlation length in our 3D nucleus model.
It is roughly estimated as the inverse of the typical longitudinal momentum of the hard partons contributing to the color charge density. 
This length may be associated with the Bjorken variable $x$, which in turn implies that the correlation length is also related to the saturation scale $Q_s(x)$.
Addressing the challenge of advancing beyond a basic Gaussian model to more precisely depict the 3D color charge density for a specific Bjorken x is important.
The JIMWLK equation offers a framework for calculating observables from the color charge density over a wide range of Bjorken variable $x$ but assumes an extremely thin spatial distribution of color charge density. 
Then, to accurately determine the spatial rapidity dependence of an observable, a conversion from momentum rapidity to spatial rapidity is essential. 
Therefore, the feasibility of constructing a spatial 3D color charge density based on the JIMWLK equation presents a complex challenge.

Finally, we point out the gap between the current $3$D glasma model and phenomenology.
To enhance the realism of the initial conditions, it would be advantageous to utilize the methodology from the $2$D IP-glasma model~\cite{IPglasma1,IPglasma2}, 
a phenomenological model of the glasma in the mid-rapidity.
This approach allow us to account for the event-by-event fluctuation of nucleon positions, the saturation scale based on the IP-sat model~\cite{IPsat1,IPsat2}, and the expansion of the system in the transverse plane, all of which are not taken into account in the current model.
For the transition of glasma into initial hydrodynamic conditions, 
separating the glasma from the Weizs\"{a}cker-Williams(WW) field is crucial. 
One proposed method is to employ new ``usual Milne coordinates'' for observation, setting the proper time to $0$ at the nucleus's rear surface, 
unlike the current usual Milne coordinates, which link proper time of 0 with the nucleus's central longitudinal position. 
This alteration would place the majority of the WW field outside the forward light cone, 
facilitating the separation from the glasma. 
Considering the energy density and pressure in the local rest frame, defined as eigenvalues of the energy-momentum tensor, 
could potentially exclude the contribution of the WW field on the EM tensor~\cite{3Dglasma7}. 
Since the EM tensor's determinant for the WW field is zero, 
energy density and pressure in the local rest frame should effectively disappear in regions heavily influenced by the WW field.
To further align the 3D glasma model with real-world phenomena, transitioning from SU(2) to SU(3) is necessary.

\vspace{6pt} 




\authorcontributions{
Conceptualization, H.M. and X.-G.H.; methodology, H.M.; 
software, H.M.; 
validation, H.M. and X.-G.H.; 
formal analysis, H.M.; 
investigation, H.M. and X.-G.H.; 
data curation, H.M.; 
writing---original draft preparation, H.M. and X.-G.H.; 
writing---review and editing, H.M. and X.-G.H.; 
funding acquisition, X.-G.H.
All authors have read and agreed to the published version of the manuscript. 
}

\funding{This research is supported by the National Natural Science Foundation of China (Grant No. 12225502, No. 12075061, and  No. 12147101), the National Key Research and Development Program of China (Grant No. 2022YFA1604900), and the Natural Science Foundation of Shanghai (Grant No. 23JC1400200).}

\institutionalreview{Not applicable.}

\informedconsent{Not applicable.}

\dataavailability{Data are available upon request.}
 


\acknowledgments{The authors acknowledge Yukawa-21 for providing the computational}

\conflictsofinterest{The authors declare no conflicts of interest.}




\noindent 

\appendixtitles{no} 
\appendixstart



\begin{adjustwidth}{-\extralength}{0cm}

\reftitle{References}




\PublishersNote{}
%


\end{adjustwidth}
\end{document}